\documentclass[aps,prl,showpacs,twocolumn,superscriptaddress]{revtex4}

\usepackage{amsmath} 
\usepackage{amssymb} 
\usepackage{bm}
\usepackage{graphicx}
\usepackage{dcolumn}

\begin{document}


\title{Renormalization of Molecular Electronic Levels at Metal-Molecule Interfaces}

\author{J. B. Neaton}
\affiliation{The Molecular Foundry, Materials Sciences
             Division, Lawrence Berkeley National Laboratory,
             Berkeley, CA 94720, USA}
\author{Mark S. Hybertsen}
\affiliation{Department of Applied Physics and Applied Mathematics and
Center for Electron Transport in Molecular Nanostructures, Columbia University, 
New York, NY 10027}
\author{Steven G. Louie}
\affiliation{The Molecular Foundry, Materials Sciences
             Division, Lawrence Berkeley National Laboratory,
             Berkeley, CA 94720, USA}
\affiliation{Department of Physics, University of California, Berkeley, CA 94720}


\begin{abstract}
The electronic structure of benzene on graphite (0001) is computed using the GW approximation for the electron self-energy. The benzene quasiparticle energy gap is predicted to be 7.2 eV on graphite, substantially reduced from its calculated gas-phase value of 10.5 eV. This decrease is caused by a change in electronic correlation energy, an effect completely absent from the corresponding Kohn-Sham gap.  For weakly-coupled molecules, this correlation energy change is seen to be well described by a surface polarization effect. A classical image potential model illustrates trends for other conjugated molecules on graphite.
\end{abstract}

\date{June 18, 2006}


\maketitle


There is renewed interest in using organic molecules as components in nanoscale electronic and optoelectronic devices \cite{ref1,ref2}, and thus a critical need has emerged for improved knowledge and control of charge transport phenomena in organic molecular assemblies \cite{ref3}.  Understanding transport across the interface between the active organic layer and the metallic electrode has proved particularly challenging, especially in the single-molecule limit. Fundamentally, charge transport is controlled in such systems by the electronic coupling of frontier molecular orbitals to extended states in the electrode, and the energetic position of these orbitals relative to the contact Fermi level.  Several recent measurements of organic thin films, self-assembled monolayers (SAMs), and single-molecule junctions have emphasized the important role of Coulomb interactions between the added hole or electron in the frontier orbitals and the metal substrate \cite{ref4,ref5,ref6,ref7,ref8,ref9,ref10,ref30}. However, most theoretical calculations of transport through organic molecules have continued to rely on some implementation of density functional theory (DFT) or semiempirical one-particle Hamiltonians \cite{ref3}.  The limitations of DFT for describing excited-state energies are well known \cite{ref11}, and implications for a DFT-based theory for nanoscale conductance have been recently discussed \cite{ref12}.

When a molecule is brought in contact with a metal, several physical effects will influence its ionization level (highest occupied molecular orbital, HOMO) and affinity level (lowest unoccupied molecular orbital, LUMO). First, the self-consistent interaction between molecule and surface will rearrange the electron density and modify the alignment of frontier orbital energies. Second, electronic coupling to extended states in the metal will further shift orbital energies and broaden discrete molecular levels into resonances.  Finally, the Coulomb interaction between the added hole or electron associated with the ionization or affinity level will result in a polarization of the metal substrate.   This additional correlation energy further stabilizes the added hole or electron, reducing the gap between affinity and ionization levels as illustrated in Fig. 1.  An accurate DFT-based approach should correctly capture the first effect \cite{ref12a}, although the use of DFT to calculate the width of resonances is under debate \cite{ref12}. Importantly however, the surface polarization response, as we show here, is completely absent from frontier orbital energies computed in DFT.

\begin{figure}[ht]
  \includegraphics[width=6.5cm]{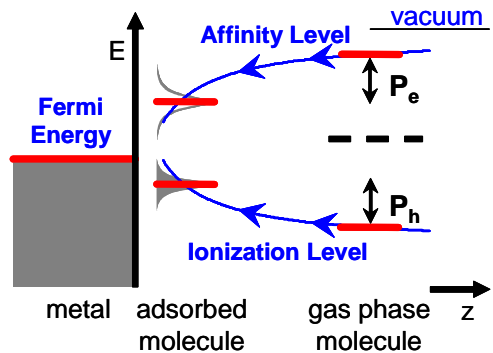}
\caption{
Schematic energy level diagram indicating polarization shifts in the frontier energy levels (ionization and affinity) of a molecule upon adsorption on a metal surface. 
}
\end{figure}

 In this Letter, we compute the electronic excited states for a clean, weakly-coupled system consisting of an aromatic molecule (benzene) physisorbed on the graphite (0001) surface. Electronic correlations are included directly within a first-principles many-electron Green function approach \cite{ref13}.  The electron self-energy is calculated from first principles within the GW approximation (GWA) \cite{ref14} using a methodology \cite{ref15} that has proved accurate for a wide range of systems \cite{ref16}.  While more generally including dynamical electronic correlation, the GWA is well known to include static, long-range image potential effects for an electron near an interface \cite{ref17}.  Using this theoretical approach, we predict a strong renormalization of the electronic gap of the benzene system (relative to its molecular gas-phase value) when it is physisorbed on a graphite (0001) surface.   The change in the electron correlation energy on adsorption can be understood as a polarization effect in this case.  An image potential model is used to illustrate trends for other aromatic molecules.  

Equilibrium geometries of molecular benzene in the gas-phase, condensed in a bulk crystalline phase, and physisorbed on graphite (0001) are determined using DFT within the local density approximation (LDA). Norm-conserving pseudopotentials \cite{ref18} are used with a plane-wave basis for the electron wavefunctions (80 Ry cutoff) for structural relaxations. The surface is modeled with a 3$\times$3 supercell containing 4 layers of graphite, a single benzene molecule, and the equivalent of 7 layers of vacuum. The theoretical in-plane bulk lattice parameter is used ($a$=2.45~\AA, $c$=6.62~\AA).  In the most stable site for adsorption, benzene rests flat on the surface centered on a three-fold site 3.25~\AA\ above a substrate carbon atom, in agreement with a previous study \cite{ref19}. For comparison, benzene is also considered in an upright position, centered above a hollow site with its closest hydrogen atom 2.21~\AA\ from the surface. Solid crystalline benzene has an orthorhombic unit cell containing four molecules (Pbca); the atomic positions within the unit cell are optimized keeping the lattice parameters $a$, $b$ and $c$ fixed to their experimental values of of 7.44, 9.55 and 6.92~\AA\ respectively \cite{ref20}.  The gas phase is modeled using a cubic supercell ($a$=13.22~\AA).  For each system, matrix elements of the self-energy operator are evaluated using a 50 Ry energy cutoff for the electronic wavefunctions, a 6 Ry cutoff for the momentum-space dielectric matrix, and a 2.9 Ry cutoff for the sum on the virtual states. This choice of parameters results in quasiparticle energy gaps converged within $\sim$~0.2 eV. 

The electron addition and removal energies of a benzene molecule in the gas-phase, calculated in the present GW approach, result in a HOMO-LUMO (quasiparticle) gap of 10.51 eV. This value agrees well with an independent GW calculation \cite{ref21}, total energy difference calculations based on DFT \cite{ref22,ref23}, and experiment \cite{ref24}. By contrast, the Kohn-Sham gap (within LDA) is 5.1 eV, substantially smaller.  
\begin{figure}[ht]
  \includegraphics[width=6.5cm]{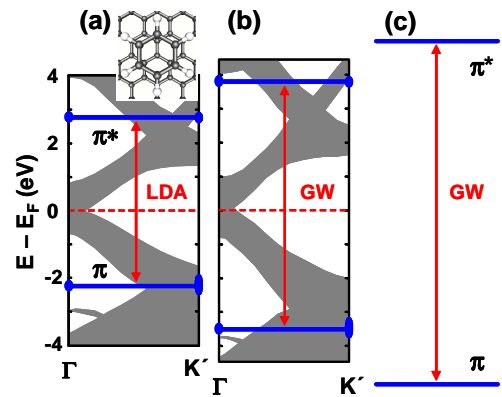}
\caption{
Calculated frontier orbital energy levels (heavy blue lines) with the indicated energy gap (red arrows) for benzene adsorbed flat on the graphite surface, plotted against the projected surface band structure of the graphite: (a) DFT (LDA) energies, (b) GW quasiparticle energies,  (c) GW quasiparticle energies of benzene in the gas phase.  Inset in (a) shows a model of the adsorption geometry.
}
\end{figure}
The electronic structure of benzene on the graphite (0001) surface along the $\Gamma$-K' direction is shown in Fig. 2. Comparing the surface-projected band structures (shaded regions) in Fig. 2(a) and 2(b), the quasiparticle bandwidth increases by about 15\% relative to LDA, in agreement with previous works \cite{ref25,ref26}. The bold horizontal lines interpolate between the benzene HOMO and LUMO states computed at $\Gamma$ and K'; the filled circles at these high-symmetry points indicate states with significant weight on the molecule. For physisorbed benzene, the Kohn-Sham (LDA) gap is 5.05 eV throughout the zone, unchanged from the corresponding LDA gas-phase value. Relative to the LDA value, the quasiparticle gap of the molecule flat on the graphite surface is much {\it larger}, 7.35 eV. However, the predicted quasiparticle gap is substantially {\it smaller} than the gas-phase value of 10.51 eV.

Table 1 summarizes the calculated HOMO-LUMO gaps of benzene in the four environments considered in this study.  Interestingly, the LDA gaps are identical for all environments. In contrast, the GW self-energy corrections exhibit noticeable variation. To understand this variation, we analyze the self energy change relative to the gas-phase.  The change $\Delta\Sigma$ for each frontier level is decomposed into Coulomb-hole ($\Delta\Sigma_{\rm CH}$), screened-exchange ($\Delta\Sigma_{\rm SX}$), and bare exchange or Fock ($\Delta\Sigma_{\rm X}$) contributions.  We find that $\Delta\Sigma_{\rm CH}$ is nearly equal for the occupied and empty frontier states, and that $\Delta\Sigma_{\rm X}$ is quite small (0.1-0.2 eV). Interestingly, the screened exchange term that is responsible for most of the difference: for the HOMO, we observe $\Delta\Sigma_{\rm SX}$~$\sim$~-2$\Delta\Sigma_{\rm CH}$, while for the LUMO $\Delta\Sigma_{\rm SX}$~$\sim$~0.  Put together, the change in correlation energy ($\Delta\Sigma_{\rm Corr}=\Delta\Sigma_{\rm CH}+\Delta\Sigma_{\rm SX}$) reported in Table 1 turns out to be nearly symmetric between the ionization and affinity levels for the benzene in each environment studied. This result is qualitatively the same as that obtained from the derivation of the image potential effect for an electron near a metal surface \cite{ref17}.

\begin{table}
\caption{\label{displacements} Benzene HOMO-LUMO gaps in the gas phase, crystal phase, and adsorbed on the graphite surface (flat and perpendicular).  First and second lines are Kohn-Sham (LDA) and quasiparticle (GW) gaps.  (For the crystal, we average over the $\pi$ and $\pi^*$ manifolds.) Third and fourth lines are calculated changes in correlation energy for the HOMO and LUMO, relative to the gas phase, determined from the full GW calculations and from an image potential model. Energies are in eV.}
\begin{center}
\begin{tabular}{l|cccc}
\hline
\hline
 & Gas & Flat & Perp & Crystal \\
 & phase & graphite & graphite & phase \\
\hline
$\Delta E_{\rm gap}$ (LDA) & 5.16 & 5.05 & 5.11 &5.07\\
$\Delta E_{\rm gap}$ (GW) & 10.51 & 7.35 & 8.10 &7.91\\
$\Delta\Sigma_{\rm Corr}$ &  &  1.45, -1.51 & 1.18, -1.17 & 1.16, -1.15\\
$\Delta\Sigma_{\rm Corr}$ (Model) &  &  1.50, -1.43 & 0.97, -0.96 & \\
\hline
\hline
\end{tabular}
\end{center}
\end{table}

To develop a more detailed model of our self energy results, we recognize that the benzene frontier orbitals are only weakly coupled to the environment.  When the orbitals have negligible overlap with the metal, the correction to the molecular self-energy operator upon adsorption depends only on the change in the screened Coulomb interaction W, i.e.            
\begin{equation}
\Delta \Sigma_{SX}({\bf r,r'}; E) = \sum_{j}^{occ} \phi_j({\bf r})\phi^*_j({\bf r'}) \Delta W ({\bf r,r'};E-E_j),
\end{equation}
where $\phi_{\rm j}$ are molecular wavefunctions and $E_j$ their eigenvalues. A corresponding expression exists for the Coulomb-hole term.   For sufficiently large metal-molecule separations, $\Delta$W is smooth and slowly-varying over the spatial extent of the molecular orbitals, and only the self-term contributes to single-particle matrix elements of Eq. (1).  Then the change in correlation energy from the surface can be reduced to 
\begin{equation}
\begin{split}
\Delta E_{\rm HOMO} &= \langle \phi_{\rm HOMO}|\Delta\Sigma_{\rm SX}+\Delta\Sigma_{\rm CH}|\phi_{\rm HOMO}\rangle \\ 
                   &\approxeq 2P_{\rm HOMO} - P_{\rm HOMO} = P_{\rm HOMO} \\
\end{split}
\end{equation}
and
\begin{equation}
\begin{split}
\Delta E_{\rm LUMO} &= \langle \phi_{\rm LUMO}|\Delta\Sigma_{\rm SX}+\Delta\Sigma_{\rm CH}|\phi_{\rm LUMO}\rangle \\
                   &\approxeq 0 - P_{\rm LUMO} = -P_{\rm LUMO}, \\
\end{split}
\end{equation}
where $P$ is the static polarization integral
\begin{equation}
P_{j} = - {1\over 2} \int\int d{\bf r}d{\bf r'} \phi_j({\bf r})\phi^*_j({\bf r'}) \Delta W ({\bf r,r'}) \phi_j({\bf r'})\phi^*_j({\bf r}).
\end{equation}
For benzene on graphite, the full GW calculations indicate that dynamical effects make a negligible contribution to $\Delta\Sigma_{\rm Corr}$, and that the self-term accounts for more than 90\% of $\Delta\Sigma_{\rm Corr}$, supporting the simplified picture of Eqs.~(2-4).

\begin{figure}[ht]
  \includegraphics[width=6cm]{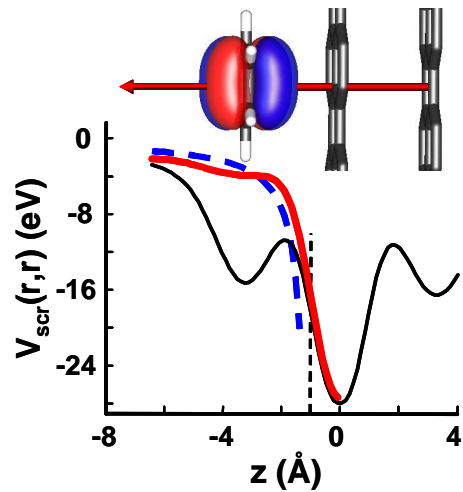}
\caption{
Static screening potential, $V_{\rm scr}({\bf r},{\bf r})$, plotted along a line through benzene adsorbed flat on graphite.  Thin, solid (black) curve is the total $V_{\rm scr}({\bf r},{\bf r})$ for the metal-molecule -system; the thick, solid (red) curve is $\Delta V_{\rm scr}({\bf r},{\bf r})$, the change upon adsorption; the heavy, dashed (blue) curve is the image potential model relative to the image plane (light, vertical dashed line).  Inset: physical model, scaled to the axis of the plot, including an isosurface plot of a frontier benzene $\pi$ orbital.
}
\end{figure}

Significant further simplification is achieved if an image potential model is sufficient for $\Delta W({\bf r},{\bf r})$.  In Fig.~3, the screening potential through the molecular adsorbate is illustrated by considering $\Delta W({\bf r},{\bf r})$ = $\Delta V_{\rm scr}({\bf r},{\bf r})$, where $V_{\rm scr}({\bf r},{\bf r})$ results from the screening response to the added electron (or hole) \cite{ref15}. The difference between the adsorbed molecule and isolated molecule, $\Delta V_{\rm scr}({\bf r},{\bf r})$, is compared with the image potential model, $\Delta V_{\rm scr}({\bf r},{\bf r})$$\rightarrow$$1/4|z-z_0|$. The image plane position z$_0$ is explicitly determined, in a separate calculation \cite{ref27}, to be 1~\AA\ beyond the last surface plane for our graphite slab.    From Fig. 3 it can be seen that, over the spatial range of the molecular orbital, an image potential captures the main effect. Using the value of the image plane position computed above and the frontier orbitals, $P_{\rm HOMO}$ and $P_{\rm LUMO}$ for benzene flat and perpendicular on the graphite surface are calculated using the image potential model.  As shown in Table 1, the image potential model is quite accurate for the flat case and captures most of the effect for the perpendicular case (within 0.2 eV). The simple image potential model neglects the internal screening response of the molecule to the polarization of the metal surface.  While small for a flat molecule oriented parallel to a surface, a significant molecular polarizability perpendicular to the metal surface leads to an additional energy gain, increasing the $P_{\rm j}$.

\begin{table*}
    \caption{\label{table:struct} For selected molecules, measured gas-phase ioniziation energies and electron affinities \cite{ref24} are combined with an image model for polarization energies to predict adsorbate HOMO-LUMO gaps for molecules flat on graphite (all in eV).
     }
    \begin{ruledtabular}
    \begin{tabular}{ccccccc}
Molecule & Expt IP & Expt EA & Gas-phase gap &  $P_{\rm HOMO}$ & $P_{\rm LUMO}$ & Adsorbate gap \\
       \tableline
  Naphthalene
   & 8.14 & -0.20 & 8.34 & 1.41 & 1.39 & 5.54\\
  Anthracene
   & 7.44 & 0.53 & 6.91 & 1.32 & 1.30 & 4.29\\
  Tetracene
   & 6.97 & 0.88 & 6.09 & 1.24 & 1.23 & 3.62\\
  Pentacene
   & 6.63 & 1.39 & 5.24 & 1.18 & 1.18 & 2.88\\
  Coronene
   & 7.29 & 0.47 & 6.82 & 1.19 & 1.17 & 4.46\\
    \end{tabular}
    \end{ruledtabular}
\end{table*}

Provided that molecular resonances are well separated from the metal Fermi energy, the polarization model for $\Delta\Sigma$ should be broadly applicable. To illustrate the impact of the change in correlation energy, we use the image model to predict the renormalized gaps for members of the acene series and coronene adsorbed flat on graphite (Table 2).  For the larger molecules in the series, the change in gap becomes dramatic, e.g. the pentacene gap is predicted to diminish by nearly a factor of two on a graphite surface. 

The role of geometry and morphology on changes in polarization energy in organic systems can be subtle \cite{ref28}, but the impact has been directly measured for organic films on metal substrates using photoemission and inverse photoemission \cite{ref6}.  Adsorbate frontier orbital energies can be probed directly by STM, provided the HOMO-LUMO gap is small enough for the resonant tunneling regime to be experimentally accessible \cite{ref6,ref29}.  From Table 2, tetracene and pentacene are within typical measurement range ($\pm$2.5 V), while anthracene and coronene are marginal. In a recent study of pentacene adsorbed on ultrathin NaCl on Cu(111) \cite{ref30}, gaps of 3.3, 4.1 and 4.4 eV are observed for NaCl thicknesses of one, two and three monolayers, respectively.  Our modeled value of 2.9 eV for direct adsorption on the graphite surface fits well with this progression.  

In conclusion, we find that the correlation contribution to the frontier molecular orbital energies depends sensitively on environment.  In the examples studied here, the change in correlation energy is dominated by a polarization effect. The impact of electrode surface polarization on spectroscopic measurements must be carefully assessed for each metal-molecule system.  For organic films or SAMs on a metal, the polarization contribution from neighboring molecules can also be quite significant. For molecular systems where the frontier orbitals have stronger electronic coupling to the metal and the resultant resonances overlap with the Fermi energy, the role of dynamical charge transfer is expected to be considerable.  Future investigations must address the nature of additional contributions to the self energy in the event of stronger coupling.

\begin{acknowledgments}
We thank Prof. G. W. Flynn and Dr. C. D. Spataru for useful discussions. This work was supported by the Director, Office of Science, Office of Basic Energy Sciences, Division of Materials Sciences and Engineering, of the U.S. Department of Energy under Contract No. DE-AC03-76SF00098, the National Science Foundation under Award No. DMR04-39768, the Nanoscale Science and Engineering Initiative of the National Science Foundation under NSF Award Number CHE-0117752, and the New York State Office of Science, Technology, and Academic Research (NYSTAR). Computational resources were provided by NERSC and NPACI.
\end{acknowledgments}


\end{document}